\documentclass[%
reprint,
twocolumns,
amsmath,
amssymb,
superscriptaddress,
aps,
floatfix,
table,
xcdraw,
nofootinbib
]{revtex4-2}
\usepackage[english]{babel}    
\usepackage[utf8]{inputenc}

\usepackage{graphicx}
\graphicspath{ {figs/} }

\usepackage{dcolumn}
\usepackage{bm}
\usepackage{graphicx,wrapfig}

\usepackage{braket}

\usepackage{amsmath,amssymb}
\usepackage{caption}
\usepackage{subcaption}
\usepackage{amsmath}
\usepackage{mathtools}
\usepackage{array}
\usepackage[]{xcolor}
\usepackage{float}

\usepackage{notes2bib}

\usepackage[colorlinks = true,
            linkcolor = blue,
            urlcolor  = blue,
            citecolor = blue,
            anchorcolor = blue]{hyperref}
            
\usepackage{cleveref}
\Crefformat{figure}{#2Fig.~#1#3}
\Crefmultiformat{figure}{Figs.~#2#1#3}{ and~#2#1#3}{, #2#1#3}{ and~#2#1#3}

\crefname{section}{Section}{Sections}
\crefname{appendix}{Appendix}{Appendices}

\definecolor{code}{HTML}{505050}
\NewDocumentCommand{\codeword}{v}{%
\texttt{\textcolor{code}{#1}}%
}

\begin{document}

\preprint{APS/123-QED}

\title{Improving and benchmarking NISQ qubit routers}

\author{Vicente Pina-Canelles}
\email{vicente.pina@meetiqm.com}
\affiliation{%
IQM Germany GmbH, Nymphenburgerstrasse 86, 80636 Munich, Germany
}
\affiliation{%
Department of Physics and Arnold Sommerfeld Center for Theoretical Physics, Ludwig-Maximilians-Universit\"at M\"unchen, Theresienstrasse 37, 80333 Munich, Germany
}%
 \author{Adrian Auer}
\affiliation{%
IQM Germany GmbH, Nymphenburgerstrasse 86, 80636 Munich, Germany
}
\author{Inés de Vega}
\affiliation{%
IQM Germany GmbH, Nymphenburgerstrasse 86, 80636 Munich, Germany
}
\affiliation{%
Department of Physics and Arnold Sommerfeld Center for Theoretical Physics, Ludwig-Maximilians-Universit\"at M\"unchen, Theresienstrasse 37, 80333 Munich, Germany
}%

 \date{\today}

\begin{abstract}

Quantum computers with a limited qubit connectivity require inserting SWAP gates for qubit routing, which increases gate execution errors and the impact of environmental noise due to an overhead in circuit depth. In this work, we benchmark various routing techniques considering random quantum circuits on one-dimensional and square lattice connectivities, employing both analytical and numerical methods. We introduce circuit fidelity as a comprehensive metric that captures the effects of SWAP and circuit depth overheads. Leveraging a novel approach based on the SABRE algorithm \cite{Li2019SABRE}, we achieve up to $84\%$ higher average circuit fidelity for large devices within the NISQ range, compared to previously existing methods. Additionally, our results highlight that the optimal routing choice critically depends on the qubit count and the hardware characteristics, including gate fidelities and coherence times.

\end{abstract}

\maketitle

\section{Introduction} \label{sec:introduction}

In digital quantum computation, a quantum circuit or algorithm has a minimum requirement in terms of the number of quantum gates it comprises, and the depth required to execute them, assuming a complete connectivity of the qubits (i.e., that an entangling gate can be executed on any pair of qubits). However, some quantum computing platforms have connectivity limitations---like architectures based on superconducting qubits, which are typically restricted to a planar surface \cite{Li2019}---requiring the implementation of SWAP gates to effectively execute gates between qubits that are not physically connected. This introduces a considerable overhead in the form of extra two-qubit gates and, therefore, typically also in the depth of the quantum circuits.

While finding the optimal placement of SWAP gates when mapping a quantum circuit to a given connectivity is NP-complete \cite{Siraichi2018}, many heuristics-based methods exist for finding sub-optimal SWAP placements within efficient runtimes \cite{Finigan2018, Siraichi2018, Zulehner2018, Li2019SABRE, qiskit2024, Cowtan2019, Zhou2020}. For NISQ devices \cite{Preskill2018}, it is crucial to choose the most effective routing method for a given quantum circuit to minimize the resources needed for circuit execution and therefore the errors introduced. To this end, in this manuscript we analyze different ubiquitous routing techniques using random quantum circuits as a benchmark.

Previous benchmarks of routing techniques typically focus only on studying the number of SWAP gates and/or the circuit depth separately \cite{Holmes2020, McKinney2022, Wintersperger2022, Weidenfeller2022, Bandic2023, Rached2024, Li2023SABRE}. In contrast, this manuscript emphasizes the importance of using circuit fidelity as a metric, which inherently accounts for the hardware performance of NISQ computers. Circuit fidelity combines both the number of SWAP gates and circuit depth, capturing any possible trade-offs in the context of noisy hardware. By adopting this holistic approach, we reveal that such trade-offs are significant and strongly influence the effectiveness of different routing methods or heuristics. Specifically, when analyzing random circuits and evaluating circuit fidelity as a function of the number of qubits averaged over an ensemble of random circuits, a crossover point emerges where the best routing scheme transitions from one technique to another. This crossover is robust across various hardware parameters but occurs at different qubit numbers, emphasizing that the choice of the best routing method is influenced not only by the qubit count but also by the critical hardware characteristics, including two-qubit gate fidelities and qubit coherence times.

\begin{figure}
    \centering
    \subfloat[\label{fig:path}]{%
		\includegraphics[width=0.45\columnwidth]{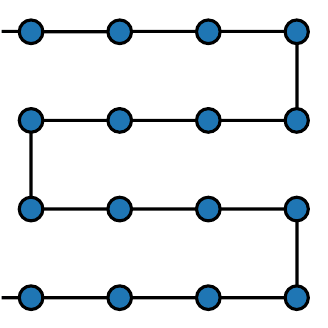} 
	}\hfill
	\subfloat[\label{fig:square}]{%
		\includegraphics[width=0.45\columnwidth]{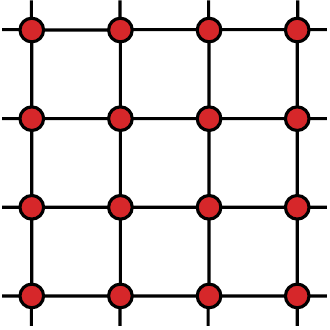} 
	}\hfill
    \caption{(a) Path graph connectivity and (b) square lattice connectivity. Each vertex represents a qubit, and each edge represents a coupling among them.}
    \label{fig:connectivities}
\end{figure}

Using the circuit fidelity as a metric, and based on our analysis, we propose the ``basic+decay" heuristic, a modification to the SABRE routing technique \cite{Li2019SABRE} that provides reductions in depth of up to $21.9\%$ on individual circuits ($3.9\%$ on average) compared to the previous best technique, while keeping the same number of SWAP gates. These reductions translate into substantial fidelity improvements, reaching up to $84\%$ for square lattice connectivity. Notably, the ``basic+decay" heuristic demonstrates superior performance, particularly beyond the discussed crossover point, making it specially effective for larger devices within the intermediate scale.

\section{Problem statement} \label{sec:problem-statement}

A quantum circuit is defined as the sequence of $G$ quantum gates, in order of application \cite{Wille2011},
\begin{equation}
    \mathcal{G}=(g_i)_{i=1}^G \, , \label{eq:quantum-circuit}
\end{equation}
where each individual gate $g_i$ can be a single-qubit gate or a multi-qubit gate, and the total number of gates is given by $G$. The depth of a quantum circuit, $D$, is defined as the minimum number of layers of gates that can be applied in parallel \cite{Nielsen2010-ft}.

To harness the full potential of a Hilbert space that grows exponentially with the number of qubits $N$ of a QPU, entanglement across all qubits is needed. This is achieved by using entangling multi-qubit operations, such as two-qubit gates (TQGs). Some physical implementations of quantum computers, like those based on trapped ions, have a complete graph connectivity, which allows for the interaction of any pair of qubits. However, other physical implementations, like those based on superconducting circuits or neutral atoms, allow only for a limited set of pairs of qubits to directly interact. The \textit{connectivity} of the device is defined as the graph $\mathcal{C} = (V, E)$, in which each vertex $V$ represents a qubit, and each edge $E$ represents a coupling between them \cite{Deb2021} (see \Cref{fig:connectivities} for two examples).

\begin{figure*}[h!btp]
	\subfloat[\label{fig:swap-path}]{%
		\includegraphics[height=.45\textwidth]{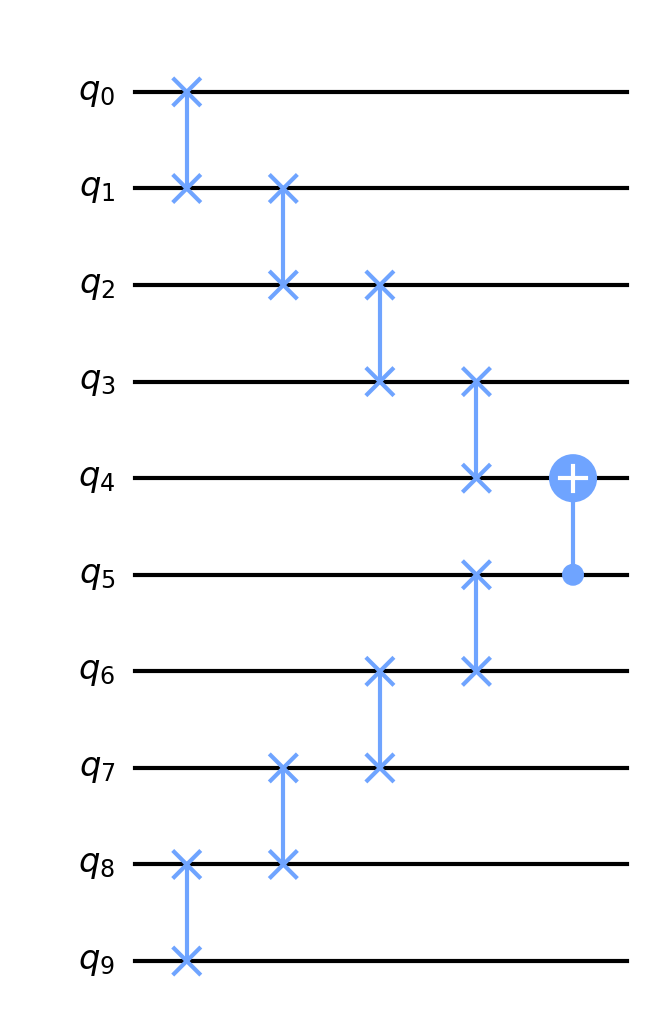} 
	}\hfill
	\subfloat[\label{fig:binary-tree-depth}]{%
		\includegraphics[height=.45\textwidth]{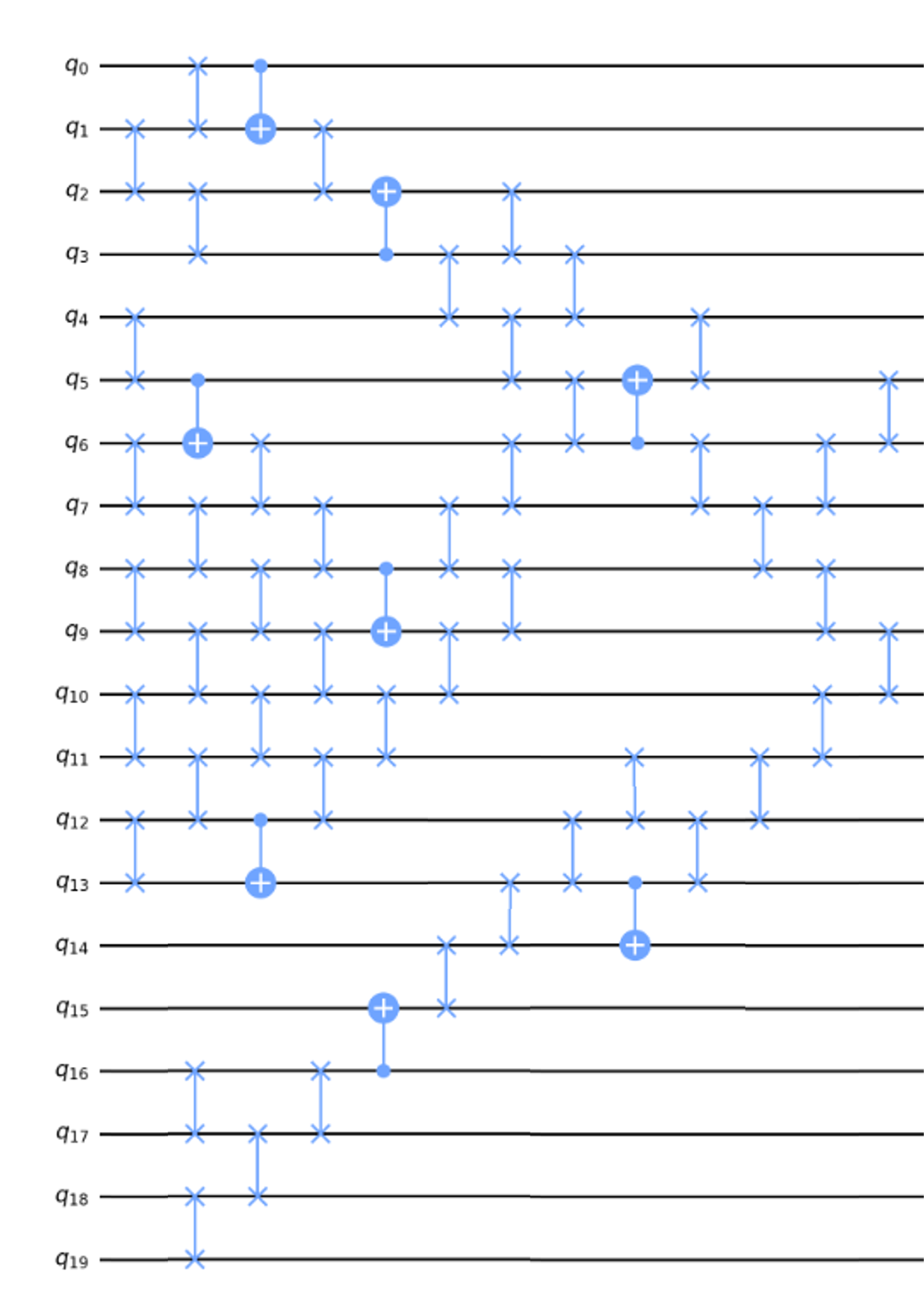} 
	}\hfill
	\subfloat[\label{fig:binary-tree-depth-2}]{%
		\includegraphics[height=.45\textwidth]{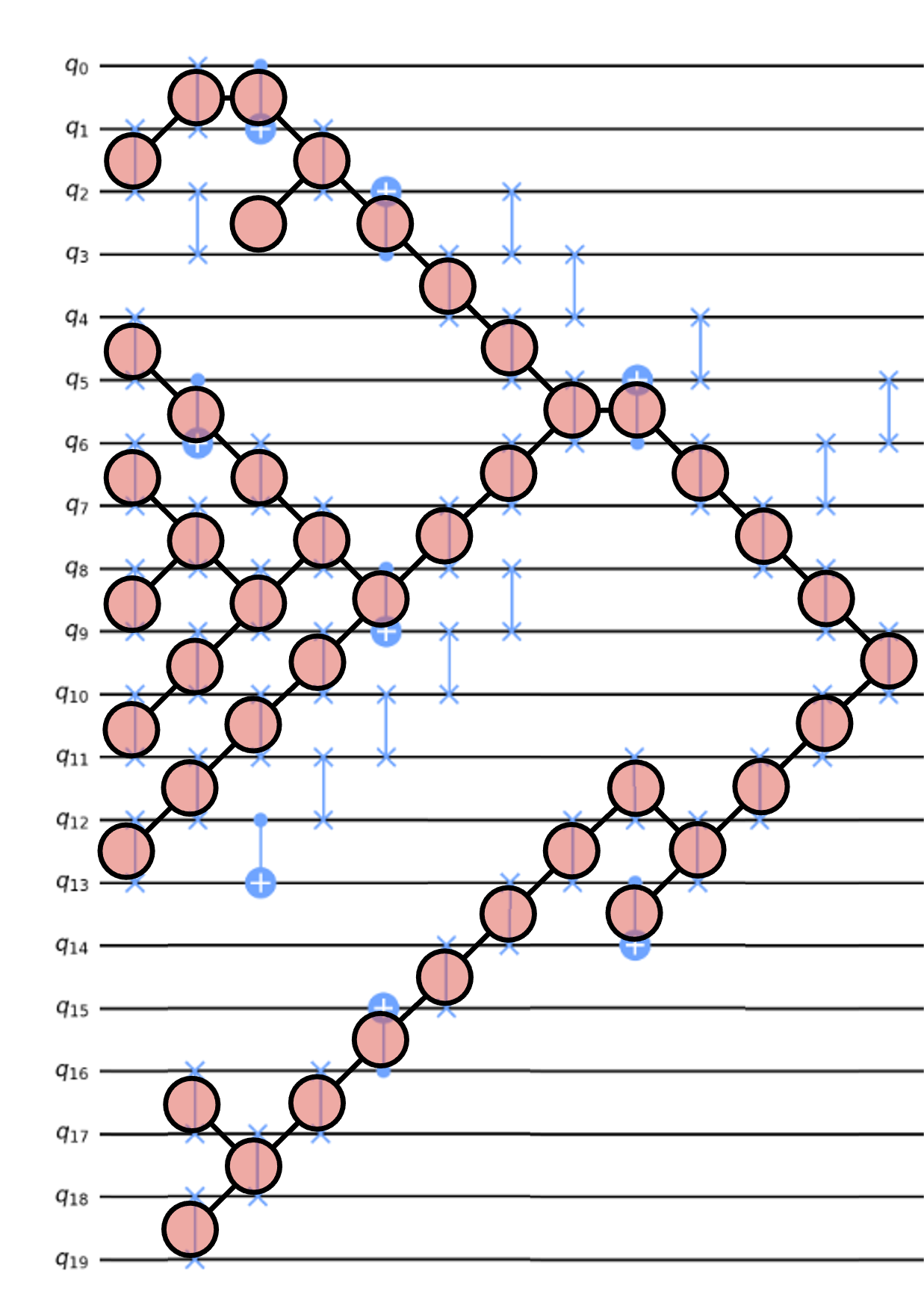} 
	}
	\caption{(a) SWAP network incurred by a single TQG. (b) SWAP network incurred by a large number of TQGs, taken from a routed random circuit, comprising $\Theta(N)$ total gates and spanning depth $\Theta(\log(N))$ given its binary tree-like structure. (c) The same SWAP network, with a binary tree superimposed, where each node corresponds to a TQG, and each edge connects a TQG to another TQG on the previous circuit layer, acting on the qubits involved in it.}
\end{figure*}

Quantum circuits generally require a complete graph connectivity. To execute these quantum circuits on a device with limited connectivity, one must implement some form of \textit{qubit routing}. Usually, this process consists on the insertion of SWAP gates, which are TQGs that act on a pair of qubits by exchanging their states. These SWAP gates are used to move the virtual qubit states around the physical qubits, in such a way that all the TQGs in $\mathcal{G}$ are acting on pairs of physical qubits that are connected at the appropriate times. This process returns a modified quantum circuit, $\tilde{\mathcal{G}}$, logically equivalent to the original one except for a final permutation of the qubit indices.

The problem of finding the optimal SWAP network for an arbitrary quantum circuit and connectivity is NP-complete \cite{Siraichi2018}. For this reason, qubit routers used for more than just a few qubits are usually based on heuristics \cite{Finigan2018, Siraichi2018, Zulehner2018, Li2019SABRE, qiskit2024, Cowtan2019, Zhou2020}, which usually provide sub-optimal routings within efficient runtimes.

An example of such a heuristics-based router is the SABRE algorithm \cite{Li2019SABRE}, publicly available as implemented in Qiskit \cite{qiskit2024}. SABRE has been previously benchmarked as the best performing across several standard routing techniques \cite{Li2023SABRE} and is regarded as state of the art \cite{LightSABRE, Kurniawan2024, Escofet2024}. The algorithm works by choosing SWAP gates one by one, according to a heuristic loss function designed to minimize the number of SWAP gates required and the depth overhead incurred. This is done by defining the ``front layer'' $F$ of gates which don't have any unexecuted predecessor gates, and an ``extended layer'' of gates $E$ which come after the front layer. The heuristic loss function is calculated for each potential SWAP gate as \cite{Li2019SABRE}
\begin{align}
    H = \Big( & \frac{1}{|F|} \sum_{\mathrm{gate} \in F} D[\pi(\mathrm{gate.q_1})] [\pi(\mathrm{gate.q_2})] \label{eq:sabre-basic-term} \\
    + W \times & \frac{1}{|E|} \sum_{\mathrm{gate} \in E} D[\pi(\mathrm{gate.q_1})] [\pi(\mathrm{gate.q_2})] \Big) \label{eq:sabre-lookahead-term} \\
    \times &\mathrm{max} \big(\mathrm{decay}(\mathrm{SWAP.q_1}), \mathrm{decay}(\mathrm{SWAP.q_2})\big) \, . \label{eq:sabre-decay-factor}
\end{align}

It is divided into the ``basic'' term \eqref{eq:sabre-basic-term}, which quantifies how many of the pairs of qubits in $F$ are brought closer together by the SWAP gate considered; the ``lookahead'' term \eqref{eq:sabre-lookahead-term}, which quantifies how many of the pairs of qubits in $E$ are brought closer together; and the ``decay'' factor \eqref{eq:sabre-decay-factor}, which deters the application of consecutive gates on the same qubit, to encourage the parallelization of gates \bibnote{$|F|$ and $|E|$ represent the number of gates in the front layer $F$ and extended layer $F$, respectively. $\mathrm{gate.q}_i$ is the $i$-th qubit involved in the gate, $D[q_1][q_2]$ represents the distance in the coupling graph between qubits $q_1, q_2$, and $\pi(q)$ is the position of qubit $q$ in the coupling graph after applying the permutation of the pertinent SWAP gate. $\mathrm{decay}(q)$ is the decay factor, which is equal to $1 + \delta$, where $\delta$ is some number $\delta > 0$ if a gate has been recently applied on qubit $q$ and $0$ otherwise. We direct the reader to Ref.~\cite{Li2019SABRE} for a more detailed explanation of the loss function and the method in general.}.

As described in Ref.~\cite{Li2019SABRE} and implemented in Qiskit \cite{qiskit2024}, the SABRE algorithm has three different heuristics the user can choose from, according to which terms of the loss function are used:
\begin{enumerate}
    \item \codeword{"basic"}: the loss function contains only the ``basic'' term \eqref{eq:sabre-basic-term}.
    \item \codeword{"lookahead"}: the loss function contains both the ``basic'' \eqref{eq:sabre-basic-term} and ``lookahead'' terms \eqref{eq:sabre-lookahead-term}.
    \item \codeword{"decay"}: the loss function contains all three terms \eqref{eq:sabre-basic-term}-\eqref{eq:sabre-decay-factor}. Throughout this manuscript, we refer to this heuristic as \textit{lookeahead+decay}.
\end{enumerate}
Additionally to the heuristics for the SABRE loss function previously described, we introduce in our analysis a novel heuristic that includes only the ``basic'' term \eqref{eq:sabre-basic-term} and the decay factor \eqref{eq:sabre-decay-factor}, which is referred as \textit{basic+decay} throughout the manuscript. 

During the finalization of this work, we became aware of related advances in Ref. \cite{LightSABRE} that offer the potential to reduce the SWAP count and/or depth when compared to previous other methods, by using the SABRE framework but with the design of two novel heuristics. These are the ``depth'' heuristic, designed to reduce depth by sacrificing routing runtime, given that it must compute the effect on the depth of the following gates after considering every SWAP; and the ``critical path'' heuristic, which aims to reduce the depth by instead identifying the gates in the critical path of the quantum circuit (those that constitute a bottleneck for depth) and giving priority to the SWAPs that enable their execution.

\section{Scaling analysis} \label{sec:scaling-analysis}

In this section, we predict the scaling properties of quantifiable metrics used to benchmark the effect of a given qubit routing on the performance of executing a quantum circuit~$\mathcal{G}$ on a connectivity~$\mathcal{C}$. We consider the different heuristics of the SABRE algorithm described in \Cref{sec:problem-statement}, and then confirm these predictions with numerical results in \Cref{sec:results}, along with benchmarks for other ubiquitous routing techniques.

The metrics we study are the following:
\begin{enumerate}
\item The \textbf{number of SWAP} gates, $S$. The execution of quantum gates introduces a significant source of error in the computation, as they are subject not only to environmental noise during their implementation, but also to imperfect control. These SWAP gates may need to be transpiled into other TQGs and single-qubit gates, which typically have error rates one order of magnitude smaller than TQGs.

\item The \textbf{depth} of the routed quantum circuit, $\tilde{D}$, which fulfills $\tilde{D} \geq D$ where $D$ is the depth of the original circuit. The depth of a quantum circuit is a measure of how much time is required to execute it. It is related to the errors introduced by environmental noise on top of those accounted for by the infidelities of the gates applied.

\item The \textbf{fidelity} of execution, which depends on both $S$ and $\tilde{D}$.
\end{enumerate}

Ideally, we wish to find properties of the connectivities and the routers which are predictive of the performance of routing any given quantum circuit. However, each quantum circuit has a different structure, given by its sequence of gates $\mathcal{G}$. These generally result in vastly different SWAP networks, potentially yielding different performances \cite{Holmes2020, McKinney2022, Hu2022, Wintersperger2022, Weidenfeller2022, Bandic2023, Rached2024}.

As a benchmark, we use random quantum circuits (sequences of TQGs placed on uniformly sampled pairs of qubits), because they constitute a family of circuits general enough to be useful, while being specific enough that it is allows us to make predictions and analyze the results. In detail, random circuits have been argued to model arbitrary quantum state preparation \cite{Cross2019, Yuan2024} and have been linked to attempts at capturing an ``average'' quantum circuit \cite{Wack2021}. Additionally, certain NISQ algorithms exhibit a similar structure, such as the Quantum Approximate Optimization Algorithm \cite{Farhi2014}.

\subsection{SWAP count} \label{sec:swap-gates}

We set out to predict the scaling of the SWAP count for the different heuristics of the SABRE algorithm mentioned in \Cref{sec:problem-statement}. For that purpose, let us assume first that the quantum circuit to be routed consists of just one TQG acting on qubits $q_j$ and $q_k$, on a device in which they are separated by a number of edges $L_{jk}$ along the shortest path between them. Then, the routing necessarily requires at least $L_{jk}-1$ SWAP gates in order to bring $q_j$ and $q_k$ to two physical qubits that are connected.

For a random circuit, at any point throughout the circuit, there is an equal probability that the next TQG will be applied on any pair of qubits. Thus, the average number of SWAP gates introduced per random TQG is, when routed naively (without taking into account the rest of TQGs), equal to $\langle L \rangle_\mathcal{C}-1$, where $\langle L \rangle_\mathcal{C}$ is the average shortest path for a given connectivity $\mathcal{C}$,
\begin{equation}
    \langle L \rangle_\mathcal{C}(N) = \frac{1}{N(N-1)/2} \sum_{k>j}^N \sum_{j=1}^N L_{jk} \, .\label{eq:L_general_1}
\end{equation}

Therefore, approximately, the total SWAP count $S$, after routing a random circuit that comprises $G \gg N$ TQGs, on connectivity $\mathcal{C}$, is given by
\begin{equation}
    S(N) \approx \big(\langle L\rangle_\mathcal{C}(N) - 1 \big) \, G \, . \label{eq:scaling-SWAP gates}
\end{equation}

However, $S$ can be reduced by considering TQGs not as gates isolated from each other, but rather in bundles, and choosing SWAP gates in such a way that they bring more than one pair of qubits closer together. This reduction can, on average, improve the scaling by a multiplicative factor. This is because a SWAP gate can only move two qubits closer to their destination if they should move in ``opposite'' directions, therefore the overlap between their paths is very small. However, it is important to note that very specific sequences of TQGs (e.g., that corresponding to the quantum circuit for the Quantum Fourier Transform \cite{Fowler2004}) could allow for a circumvention of this linear-order scaling.

We find that the explicit dependence of $\langle L \rangle_\mathcal{C}$ with $N$ for lattice graph connectivities is
\begin{equation}
\langle L \rangle_\mathcal{C}(N) = a_\mathcal{C} \, N^{1/d} \label{eq:L_as_N} \, ,
\end{equation}
where $a_\mathcal{C}$ is a constant specific for each connectivity, and $d$ is the dimension of the lattice, which is the number of dimensions it grows in when adding more qubits (e.g., $d=1$ for the path connectivity and $d=2$ for the square connectivity). We provide the derivation of \eqref{eq:L_as_N} for the path and square connectivities in \Cref{app:L_as_N}.

\subsection{Depth} \label{sec:depth}

While a random quantum circuit can be executed in depth $\Theta(\frac{G}{N})$ on a complete graph connectivity \cite{Yuan2024, Cross2019}, the depth overhead incurred by routing to a limited connectivity is considerably higher.

While analytically studying the scaling of depth is difficult because of the heuristic nature of practical routing methods, in this subsection we focus our efforts in explaining the difference in scaling between the \textit{basic} and \textit{lookahead} SABRE heuristics.

\begin{figure*}[h!btp]
	\includegraphics[width=.95\textwidth]{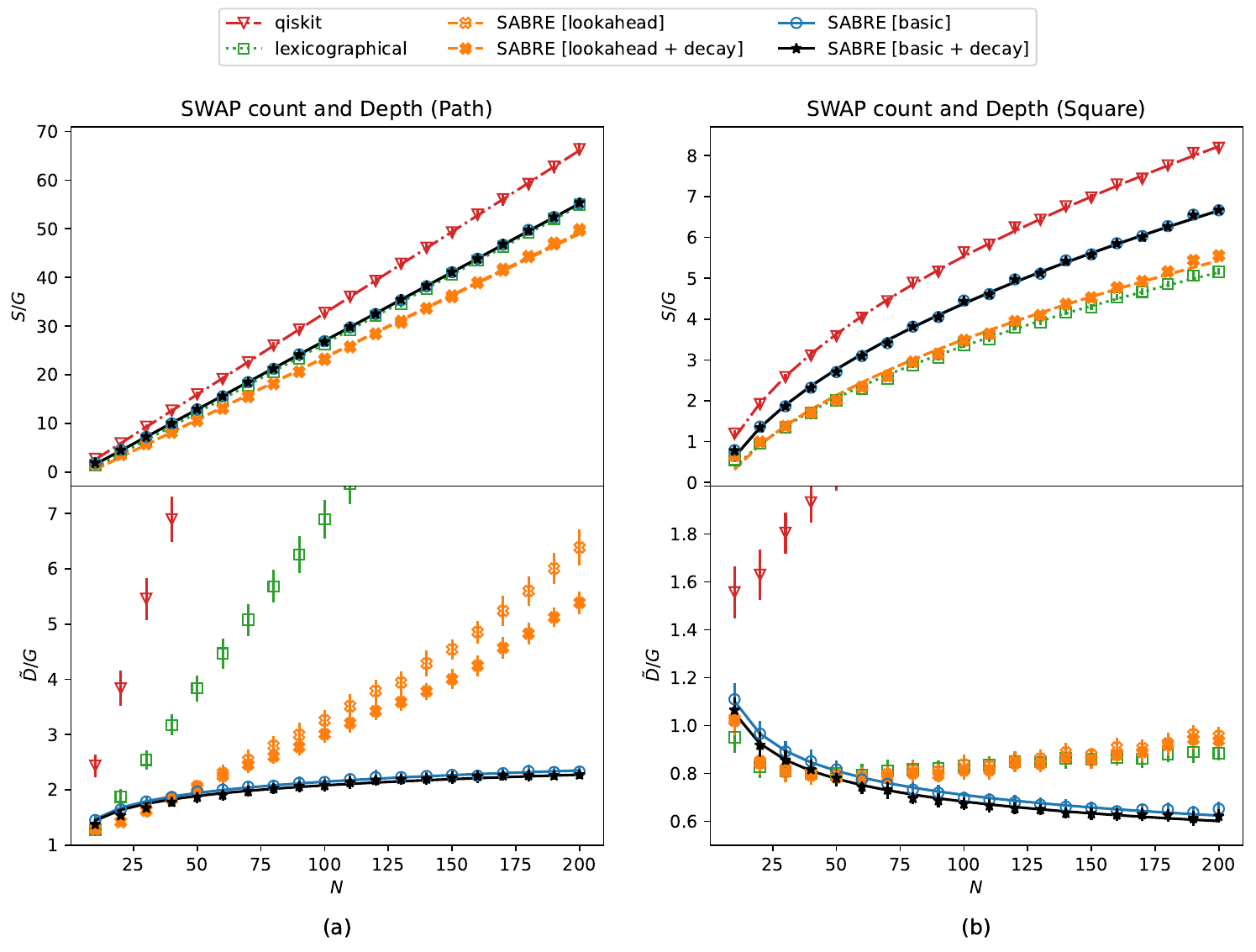} 
 {\phantomsubcaption\label{fig:combined-path}
  \phantomsubcaption\label{fig:combined-square}}
	\caption{Average SWAP gate count $S$ (top) and average depth $\tilde{D}$ (bottom) ratios with respect to the number of gates of the original random circuits, $G$, over $50$ random circuits, for (a) the path connectivity, where the SWAP count has been fit to a curve of the form \eqref{eq:swap-curve} (setting $d=1$) and the depth incurred by the \textit{basic} SABRE heuristics has been fit to a curve of the form \eqref{eq:curve-depth-1d}; and (b) the square connectivity, where the SWAP count has been fit to a curve of the form \eqref{eq:swap-curve} (setting $d=2$) and the depth incurred by the \textit{basic} SABRE heuristics has been fit to a curve of the form \eqref{eq:curve-depth-2d}. Error bars represent standard deviation. The SWAP counts for the \textit{basic} and \textit{basic+decay} heuristics overlap, and so do the ones for the \textit{lookahead} and \textit{lookahead+decay} heuristics.} \label{fig:results}
\end{figure*}
tp
\begin{figure*}[h!btp]
	\includegraphics[width=.95\textwidth]{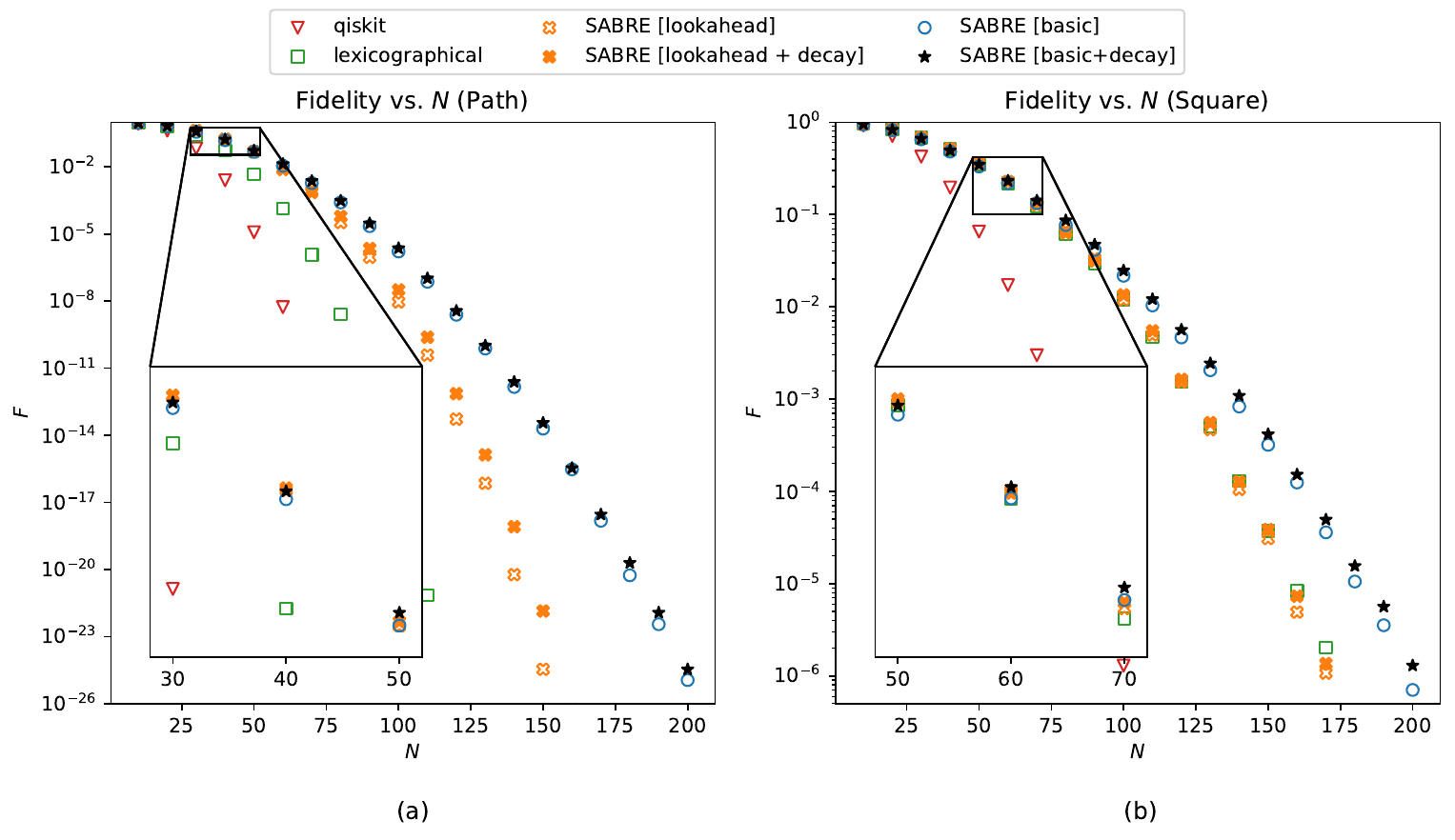} 
    {\phantomsubcaption\label{fig:fidelity-path}
    \phantomsubcaption\label{fig:fidelity-square}}
	\caption{Average quantum circuit execution fidelity according to Eq.~\eqref{eq:fidelity} over $50$ random circuits, for (a) the path connectivity, for which the \textit{basic+decay} heuristic outperforms other techniques for $N \geq 40$, and (b) the square connectivity, for which the \textit{basic+decay} heuristic outperforms other techniques for $N \geq 60$.} \label{fig:fidelity}
\end{figure*}

\subsubsection{Path connectivity} \label{sec:depth-1d-lattices}

We use the Path connectivity as a starting point for our analysis, because there is only one possible path along its edges between any two qubits $q_j$ and $q_k$, which allows us to study the scaling of depth in a simple way.

\paragraph{Basic heuristic---} We model a routed random circuit on a Path connectivity using the \textit{basic} SABRE heuristics as follows:
\begin{enumerate}
    \item Because each layer of TQGs of a random circuit contains $\Theta(N)$ gates \cite{Yuan2024}, the front layer $F$ also contains $\Theta(N)$ gates.
    \item A constant fraction of the gates in $F$ act on pairs of qubits that are $\Theta(N)$ qubits apart \cite{Yuan2024}.
    \item Each SWAP path introduced for a gate between qubits $q_j$ and $q_k$ has a shape like that of \Cref{fig:swap-path}, containing $L_{jk}-1$ SWAP gates and spanning $\lceil L_{jk}/2 \rceil$ circuit layers.
    \item Naturally, some of the ``smaller'' paths separated by $\Theta(1)$ qubits fit into ``larger'' ones separated by $\Theta(N)$ qubits, which gives rise to binary tree-like structures when maximally parallelized (see Figures \labelcref{fig:binary-tree-depth} and \labelcref{fig:binary-tree-depth-2}). More specifically, because the paths are symmetric (see \Cref{fig:swap-path}), they give rise to balanced binary trees \cite{Black2020}, whose height (i.e., the depth of the quantum circuit slice containing the binary tree) is logarithmic with the number of gates it contains.
\end{enumerate}

We have established that the depth introduced by each binary tree-like SWAP network is logarithmic with the number of gates it contains, therefore each SWAP gate introduces a depth $\Theta\left(\frac{\log(N)}{N}\right)$ on average. From \Cref{sec:swap-gates}, we know that $\Theta(N)$ SWAP gates are introduced in one-dimensional connectivities, thus
\begin{equation}
    \Tilde{D}/G = \Theta(\log(N)) \, . \label{eq:depth-1d}
\end{equation}

\paragraph{Lookahead heuristic---} On the other hand, by including the ``lookahead'' term \eqref{eq:sabre-lookahead-term}, the consideration of both the front and extended layers $F, E$ allows the router to choose SWAP gates that would belong to two different binary trees in an arbitrary order. The emergent binary tree structures coming from considering only gates in $F$ are eliminated, consequently lacking the logarithmic scaling of the depth. Therefore, each SWAP gate must be introducing a depth $\mathcal{O}(\mathrm{poly}(N))$ on average instead, which in turn makes the \textit{lookahead} heuristics entail a scaling of depth of $\mathcal{O}(\mathrm{poly}(N))$, too.

\paragraph{Decay factor---} In both the \textit{basic} and \textit{lookahead} cases, introducing the decay factor decreases the depth because it doesn't introduce any first order effects that change their scalings. Therefore, the novel \textit{basic+decay} heuristic is predicted to perform the best in terms of depth.

\subsubsection{Square connectivity} \label{sec:depth-2d-lattices}

\paragraph{Basic heuristic---} Obtaining the scaling of depth in two-dimensional lattices may not seem so straightforward, because the SWAP network for a TQG is not forced to just one possible path across the device, like they are in the path connectivity. In fact, when considering a square lattice, the number of different possible shortest paths between two qubits $q_{aj}$ and $q_{bk}$, where $a, b$ are their horizontal coordinates, and $j, k$ are their vertical coordinates on the lattice, is
\begin{equation}
    \frac{\big(|a-b| + |j-k| \big)!}{\big(|a-b|\big)! \big(|j-k|\big) !} \, .
\end{equation}

However, the square lattice graph has only double the edges than the path graph.

The router's priority is reducing the number of gates introduced. This prompts it to choose SWAP gates that bring two or more pairs of qubits closer together. Because of this, the depth of each front layer $F$ is given by some limiting SWAP path between two distant qubits, similar to that in \Cref{fig:swap-path}. While in the path graph, the distance between the farthest qubits was $\Theta(N)$, for the square lattice it is $\Theta(\sqrt{N})$---however, this is irrelevant for the logarithmic scaling because $\log(N^p) = p \log(N)$. Therefore, we can expect that the main feature of the circuits in the path graph, i.e. the symmetric binary-tree structures, remains in the square lattice for the \textit{basic} heuristic: each SWAP gate entails on average a depth $\Theta\left(\frac{\log(N)}{N}\right)$, and $\Theta(\sqrt{N})$ SWAP gates are introduced, so we predict the depth of the \textit{basic} heuristics to scale like
\begin{equation}
    \Tilde{D}/G = \Theta\left(\frac{\log(N)}{\sqrt{N}}\right) \, . \label{eq:depth-2d}
\end{equation}

\paragraph{Lookahead heuristic---} On the other hand, similarly to what happens in the case of the path connectivity, considering gates from both $F$ and $E$ destroys the emerging binary tree structures because SWAPs that would otherwise correspond to different binary trees are chosen in an arbitrary order. While this is not necessarily true for small devices, because $F$ and $E$ contain a small number of gates each, the depth is expected to grow monotonically with $N$ for large enough devices, with a scaling $\mathcal{O}(\mathrm{poly}(N))$.

\paragraph{Decay factor---} Similarly to the case of the path graph, in both the \textit{basic} and \textit{lookahead} cases, the decay factor is expected to decrease the depth.

\subsection{Fidelity} \label{sec:fidelity}

We calculate the total fidelity of execution of a quantum circuit by assuming that each gate introduces an independent error \cite{Arute2019} and that idling qubits are affected by thermal relaxation modeled by a simple Markovian noise model,
\begin{equation}
    F = f^{\tilde{G}} \times \prod_{q=1}^N e^{-t_\mathrm{idle}^q/T_1} \, , \label{eq:fidelity}
\end{equation}
where $f$ is the fidelity of TQGs, $\tilde{G}$ is the total number of TQGs including the SWAP gates, $t_\mathrm{idle}^q$ is the idling time during which no gates are executed on qubit $q$, and $T_1$ is the thermal relaxation time.

Specifically, according to the scalings described above, the fidelity for the \textit{basic} heuristics of SABRE is
\begin{equation}
    F_\mathrm{basic}^{\mathcal{P}, \mathcal{S}} = f^{\Theta(GN^{1/d})} \times e^{-\Theta(GN^{1/d}\log(N))} \, , \label{eq:fidelity-explicit-basic}
\end{equation}
where the superindex $\mathcal{P}, \mathcal{S}$ indicates that the fidelity is that of the Path and Square connectivities, respectively, whereas the fidelity for the \textit{lookahead} heuristics is
\begin{equation}
    F_\mathrm{lookahead}^\mathcal{P} = f^{\Theta(GN)} \times e^{-\mathcal{O}(G \, \mathrm{poly}(N))} \label{eq:fidelity-explicit-lookahead-1d}
\end{equation}
for the Path graph, and
\begin{equation}
    F_\mathrm{lookahead}^\mathcal{S} = f^{\Theta(GN^{1/2})} \times e^{-\mathcal{O}(G \, \mathrm{poly}(N))} \label{eq:fidelity-explicit-lookahead-2d}
\end{equation}
for the Square graph. The main source of error for both the \textit{basic} and \textit{lookahead} heuristics is the environmental noise introduced by the big depth overhead, which is considerably smaller in the \textit{basic} case. Therefore, we predict the \textit{basic} heuristics to provide a significant increase in fidelity for large enough devices.

\section{Numerical results}

\subsection{Methods}

We performed numerical simulations to obtain data for devices comprising $10, 20, ..., 200$ qubits, which is a typical range for NISQ devices \cite{Preskill2018, Bharti2022}. For each data point, we execute $50$ different random circuits with $G=10 \times N$, and route them to devices with one-dimensional path graph and two-dimensional square lattice connectivities (see \Cref{fig:connectivities}). The construction of path connectivity graphs of $N$ qubits is trivial; for the case of square connectivity graphs, we begin by creating an $l \times l$ square grid, with $l = \lceil {N}^{1/2} \rceil$, and remove the pairs of qubits that are the furthest distance apart from each other until only $N$ qubits remain.

We study the SWAP and depth overheads introduced by routing using the different heuristics of SABRE \cite{Li2019SABRE} as implemented in Qiskit \textit{v1.2.0} \cite{qiskit2024, LightSABRE}, as well as our custom \textit{basic+decay} heuristic (see \Cref{sec:problem-statement}), the Lexicographical Comparison (LC) \cite{Cowtan2019} approach as implemented in t$|$ket$\rangle$ (pytket) \textit{v1.31.1} \cite{Sivarajah2020}, and the Qiskit \codeword{BasicSwap} transpiler pass \cite{qiskit2024} (not to be confused with the \textit{basic} heuristic of SABRE). For all of the routing techniques, we apply a random initial mapping of the virtual qubits to the physical qubits, as we are most interested in the scalings with the number of qubits and, while an appropriate qubit mapping can improve the routing \cite{Cowtan2019, Li2019SABRE}, it does so only as a smaller effect \cite{Cowtan2019} when compared to the scalings described in \Cref{sec:scaling-analysis}.

Finally, we compute the total fidelity of execution of a quantum circuit using Eq.~\eqref{eq:fidelity}, assuming $f = 99.99\%$, $t_\mathrm{TQG}=35 \mathrm{ns}$ and $T_1 = 700 \mathrm{\mu s}$, which are projected values for future NISQ devices \cite{Ezratty2023, Ghosh2013} (currently sitting around $f \approx 99.5\%$ \cite{Arute2019, Google2024}). We study a wider range of such parameters in \Cref{subsec:crossover}.

\subsection{Results} \label{sec:results}

In \Cref{fig:results}, we plot the number of SWAP gates $S$ and total depth $\tilde{D}$ resulting from routing, normalized by the number of gates of the original quantum circuit, $G$, for the path and square connectivities. In \Cref{fig:fidelity}, we plot the fidelity resulting from the execution of the routed circuits, as calculated via Eq.~\eqref{eq:fidelity}.

All the parameters coming from the fit of a curve to the data in this section are found in detail in \Cref{app:curve-fits}, and details on the dependence of $S$ and $\tilde{D}$ on the number of gates $G$ can be found in \Cref{app:g}.

\subsubsection{Path connectivity}

In \Cref{fig:combined-path}, we observe that the number of SWAP gates grows linearly with $N$, as predicted in Eq.~\eqref{eq:scaling-SWAP gates}, and confirmed by fitting the data for each router to a curve of the form
\begin{equation}
    S(N)/G = A N^{1/d} + C \, , \label{eq:swap-curve}
\end{equation}
where $A, C$ are parameters of the fit, and $d=1$.

The difference among the various routing techniques is, at most, that of a multiplicative factor. Ordered from best to worst, we have the \textit{lookahead} SABRE (with and without the decay factor), the LC approach, followed closely by the \textit{basic} SABRE (with and without the decay factor), and, finally, the \codeword{BasicSwap} transpiler pass.

Regarding the depth of the routed circuit, we observe a quick linear growth with the number of qubits for the Qiskit transpiler, \textit{lookahead} SABRE and LC. However, the \textit{basic} SABRE heuristics entail a logarithmic scaling with the number of qubits, as predicted by Eq.~\eqref{eq:depth-1d}. This scaling is confirmed by fitting the data to a curve of the form
\begin{equation}
    \tilde{D}(N)/G = A \log(N) + (1-A \log(2)) \, , \label{eq:curve-depth-1d}
\end{equation}
where $A$ is the only parameter of the fit. See \Cref{app:depth-curve} for more details on the form of this curve.

The exponential reduction of depth leads to the \textit{basic} heuristics outperforming \textit{lookahead+decay} for $N \geq 50$ (basic) and $N \geq 40$ (basic+decay). On top of the effect that this has on the overall fidelity, it also means that the time it takes to execute the quantum circuit is exponentially faster, therefore significantly reducing the time to solution.

The \textit{basic+decay} heuristic incurs an average reduction of $3.5\%$ compared to its \textit{basic} counterpart, with the best individual case yielding a reduction of $25.4\%$.

The resulting fidelity is represented in \Cref{fig:fidelity-path}, where we observe that the errors due to the depth soon take over, in such a way that the \textit{basic+decay} heuristic offers the best results for $N>40$. It is clear to see the improvement provided by the newly proposed heuristic when compared to the previous best in \Cref{fig:fidelity-improvement}, where increases of up to $256\%$ in fidelity are achieved within the range studied.

\subsubsection{Square connectivity}

In the case of the square connectivity, we observe in \Cref{fig:combined-square} a scaling of $S$ proportional to $\sqrt{N}$, as predicted in Eq.~\eqref{eq:scaling-SWAP gates}, and confirmed by fitting the data to curves of the form \eqref{eq:swap-curve}, with $d=2$. Ordered from best to worst, we have the LC approach, the \textit{lookahead} SABRE heuristics, the \textit{basic} SABRE heuristics and the \codeword{BasicSwap} transpilation pass.

On the other hand, the depth incurred by the \codeword{BasicSwap} pass increases linearly with $N$. For the LC method and and \textit{lookahead} SABRE heuristics, it decrease up to a certain threshold in the number of qubits, to then start slowly increasing again. The depth incurred by the \textit{basic} SABRE heuristics, however, decreases monotonically with the number of qubits, as predicted by Eq.~\eqref{eq:depth-2d} and confirmed by fitting the data of the \textit{basic} SABRE heuristics to a curve of the form
\begin{equation}
    \tilde{D}(N)/G = A \frac{\log(N)}{\sqrt{N}} + C \, \label{eq:curve-depth-2d}
\end{equation}
where $A, C$ are parameters of the fit.

In this case, we also find that this difference in scaling leads to the \textit{basic+decay} SABRE yielding the best depth for $N \geq 50$. It entails an average reduction in depth of $3.9\%$ when compared to the \textit{basic} heuristic, and of up to $21.9\%$ in the best individual case.

\begin{figure}
    \centering
    \includegraphics[width=0.95\columnwidth]{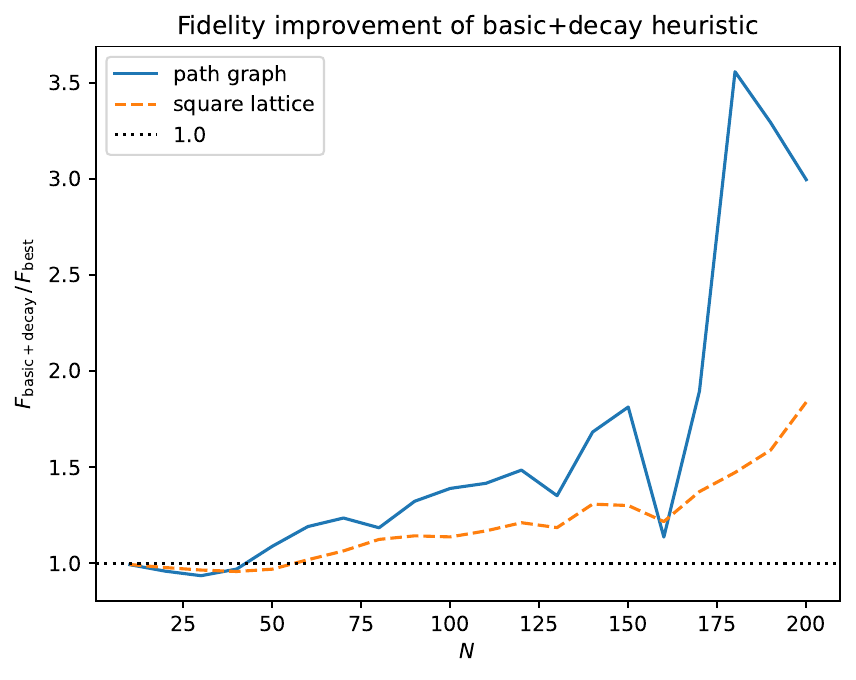} 
	\hfill
    \caption{Ratio of the quantum circuit execution fidelity when routing quantum circuits using the \textit{basic+decay} heuristic over the previous best method at each data point, for the path and square connectivities, yielding maximum improvements in fidelity of $256\%$ and $84\%$ respectively.}
    \label{fig:fidelity-improvement}
\end{figure}

As was the case for the path connectivity, we observe that the decrease in fidelity as shown in \Cref{fig:fidelity-square} is dominated by the environmental noise. Therefore, the \textit{basic} SABRE heuristics, specially \textit{basic+decay}, offer the best overall quantum circuit fidelities for devices over $60$ qubits for the parameter values chosen, potentially improving it by orders of magnitude within the range studied. The improvement compared to the previous best method at each data point is shown in \Cref{fig:fidelity-improvement}, where increases in fidelity of up to $84\%$ can be seen in the range studied.

\subsection{Dependence of the best routing technique on noise parameters} \label{subsec:crossover}

As seen in \Cref{fig:fidelity}, for the specific noise parameters chosen in our calculations, the crossover point at which the \textit{basic+decay} heuristic becomes the best technique is at $N=40$ and $N=60$ for the path and square connectivities, respectively. Such a crossover is expected to happen eventually for any such noise parameters, given that the leading terms contributing to infidelity scale exponentially better for the \textit{basic} heuristics than for the \textit{lookahead} ones (see \Cref{sec:fidelity}).

\begin{figure}
    \centering
    \includegraphics[width=0.95\columnwidth]{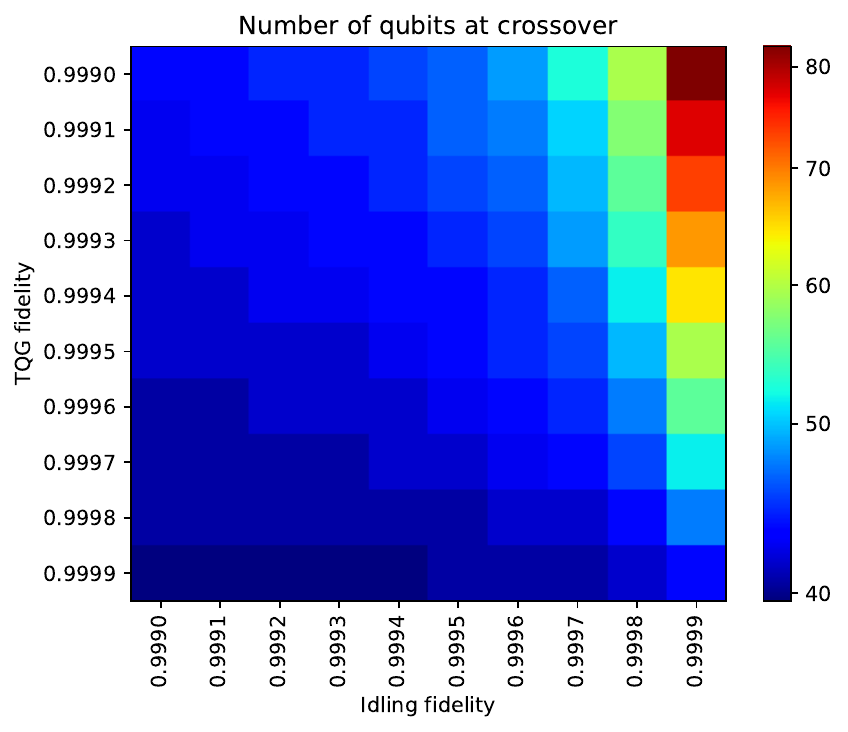} 
	\hfill
    \caption{Number of qubits at which the \textit{basic+decay} heuristic outperforms the \textit{lookahead+decay} heuristic in terms of fidelity, as a function of the TQG fidelity (vertical axis) and the idling fidelity (horizontal axis), after fitting the data to a curve of the form \eqref{eq:fidelity-fit}.}
    \label{fig:crossover}
\end{figure}

We study the dependence of the number of qubits at which such a crossover occurs with the noise parameters of the device (namely, the two-qubit gate fidelities $f$ and gate-time to decoherence time ratio $t_\mathrm{gate}/T_1$), focusing on the case of the square connectivity. In order to do so, we assign a fidelity to an ``idling'' gate, $f_\mathrm{idling} = \exp(-t_\mathrm{gate}/T_1)$. We then compute the fidelities for the \textit{lookahead+decay} and \textit{basic+decay} SABRE heuristics for different values of the gate and idling fidelities, and fit that data to a curve of the form
\begin{equation}
    F(N) = A \, B^{C N^D} \, , \label{eq:fidelity-fit}
\end{equation}
where $A, B, C, D$ are parameters of the fit. This curve is an approximation of fidelity, where the different sources of error are grouped into one that scales as $C N^D$, with each occurence of such an error introducing an infidelity $B < 1$. By doing so, we can precisely pinpoint the specific number of qubits at which such a crossover happens. We plot the results in \Cref{fig:crossover}. Which method constitutes the best is highly dependent on the noise parameters of the device, ranging from $N=40$ to $N=82$ for the studied ranges, and for the parameter values chosen. Specifically, the crossover point is larger for larger $f_\mathrm{idling}/f_\mathrm{TQG}$ ratios, as the TQG fidelity terms in Eqs.~\eqref{eq:fidelity-explicit-basic}~and~\eqref{eq:fidelity-explicit-lookahead-2d} dominate, therefore delaying the inevitable crossover coming from the depth terms.

\section{Conclusions}

In this manuscript, we benchmark a range of widely used qubit routing methods for NISQ devices and  identify the most effective ones. Our analysis, which combines analytical and numerical approaches, emphasizes circuit fidelity as a comprehensive metric that captures the performance of NISQ hardware by integrating the effects of both gate errors and overhead in circuit depth. Building on the SABRE routing technique, we identify the components of its loss functions that lead to a better performance. While the original work presenting SABRE \cite{Li2019SABRE} treated the ``basic" heuristic as merely a foundation for the more complex "lookahead+decay" heuristic, our findings demonstrate that the ``basic" heuristic is, in fact, a more effective choice for larger devices within the NISQ range. This insight enables us to design a novel loss function, ``basic+decay", achieving consistent fidelity improvements for quantum circuit execution---up to $84\%$ for the square lattice connectivity with typical NISQ parameters.

Our analysis also shows that the best choice of the routing method is highly dependent on the number of qubits, fidelity of the two-qubit gates and coherence times of the qubits. This comes from the fact that each routing method entails a different scaling of the sources of error, so that significant trade-offs arise.

The novel \textit{basic+decay} heuristic is expected to outperform on average all others studied for large enough devices ($N>60$ for the parameter values chosen). However, its performance can vary significantly between circuits. Even within the regime of large NISQ devices, results range from reductions of up to $21.9\%$ in some circuits to increases of up to $24.5\%$ in others, when applied to the square lattice. Therefore, the best strategy is to route the quantum circuit using different methods and choosing the best one on a case-by-case basis, given that quantum resources are much more costly than classical, and all these methods are efficient in their runtimes.

While this manuscript focuses on the role of router choice in the performance, the impact of a specific qubit connectivity on such performance is very relevant and has been previously studied, albeit for a limited set of algorithms and  specific connectivities in Refs. \cite{Holmes2020, McKinney2022, Hu2022, Wintersperger2022, Weidenfeller2022, Bandic2023, Rached2024}. However, a comprehensive analysis using random circuits as a benchmark for ``average'' quantum circuits on arbitrary connectivities is, to the best of our knowledge, unexplored. In the NISQ era, understanding the impact of qubit connectivity on the circuit fidelity is critical, and addressing this gap will be the  focus of our future work.

\begin{acknowledgments}
We would like to acknowledge the support of our colleagues in IQM, and specially thank Arianne Meijer.
\end{acknowledgments}

\bibliography{bibliography}

\appendix

\section{$\langle L \rangle$ for the path and square connectivities}\label{app:L_as_N}

\subsection{Path connectivity}

Assume a device with path graph connectivity $\mathcal{P}$ (see \Cref{fig:path}) comprising $N$ qubits. Then $N-1$ of its pairs of qubits have distance $1$, $N-2$ pairs of qubits have distance $2$... and $i$ pairs of qubits have distance $N-i$. Therefore, the sum of all shortest paths is $\sum_{k>j}^N \sum_{j=1}^N L_{jk} = \sum_{j=1}^N j\times (N-j) = (N-1) N (N+1)/6$. Thus, according to \eqref{eq:L_general_1}, the average shortest path length over all pairs of qubits is $\langle L \rangle_\mathcal{P} = (N+1)/3$.

\begin{table*}[t]
\begin{minipage}{.5\textwidth}
\captionof{table}{Fit parameters and $r^2$ for the SWAP count curve \eqref{eq:swap-curve} for the path connectivity. \label{tab:swap-fit-path}}
\begin{tabular}{r|c|c|c|}
\cline{2-4}
                                      & $A$   & $C$    & $r^2$   \\ \hline
\multicolumn{1}{|r|}{basic}           & 0.282 & -1.18 & 0.99996 \\ \hline
\multicolumn{1}{|r|}{basic+decay}     & 0.282 & -1.22 & 0.99996 \\ \hline
\multicolumn{1}{|r|}{lookahead}       & 0.255 & -2.00 & 0.99931 \\ \hline
\multicolumn{1}{|r|}{lookahead+decay} & 0.257 & -2.04 & 0.99943 \\ \hline
\multicolumn{1}{|r|}{lexicographic}   & 0.283 & -1.93 & 0.99987 \\ \hline
\multicolumn{1}{|r|}{qiskit}          & 0.334 & -0.755 & 0.99997 \\ \hline
\end{tabular}
\captionof{table}{Fit parameters and $r^2$ for the depth curve \eqref{eq:curve-depth-1d} for the path connectivity. \label{tab:depth-fit-path}}
\begin{tabular}{r|c|c|}
\cline{2-3}
                                  & $A$   & $r^2$ \\ \hline
\multicolumn{1}{|r|}{basic}       & 0.293 & 0.9955 \\ \hline
\multicolumn{1}{|r|}{basic+decay} & 0.276 & 0.9747 \\ \hline
\end{tabular}
\end{minipage}%
\hfill
\begin{minipage}{0.5\textwidth}
\captionof{table}{Fit parameters and $r^2$ for the SWAP count curve \eqref{eq:swap-curve} for the square connectivity. \label{tab:swap-fit-square}}
\begin{tabular}{r|c|c|c|}
\cline{2-4}
                                      & $A$   & $C$    & $r^2$  \\ \hline
\multicolumn{1}{|r|}{basic}           & 0.549 & -1.10 & 0.9988 \\ \hline
\multicolumn{1}{|r|}{basic+decay}     & 0.549 & -1.11 & 0.9989 \\ \hline
\multicolumn{1}{|r|}{lookahead}       & 0.466 & -1.16 & 0.9941 \\ \hline
\multicolumn{1}{|r|}{lookahead+decay} & 0.466 & -1.16 & 0.9943 \\ \hline
\multicolumn{1}{|r|}{lexicographic}   & 0.435 & -1.00 & 0.9980 \\ \hline
\multicolumn{1}{|r|}{qiskit}          & 0.650 & -0.961 & 0.9994 \\ \hline
\end{tabular}
\captionof{table}{Fit parameters and $r^2$ for the depth curve \eqref{eq:curve-depth-2d} for the square connectivity. \label{tab:depth-fit-square}}
\begin{tabular}{r|c|c|c|}
\cline{2-4}
                                  & $A$  & $C$   & $r^2$ \\ \hline
\multicolumn{1}{|r|}{basic}       & 1.22 & 0.152 & 0.9664 \\ \hline
\multicolumn{1}{|r|}{basic+decay} & 1.15 & 0.156 & 0.9643 \\ \hline
\end{tabular}
\end{minipage}%
\end{table*}

\subsection{Square lattice}

Recall Eq.~\eqref{eq:L_general_1} for calculating $\langle L \rangle_\mathcal{C}$, where we average the distance across all qubit pairs, considering each qubit pair only once. An alternative way to do so is to consider each pair twice, and divide by $2$:
\begin{equation}
        \langle L \rangle_\mathcal{C}(N) = \frac{1}{N(N-1)} \sum_{k=1}^N \sum_{j=1}^N L_{jk} \label{eq:L_general_2}\, .
\end{equation}

Assume a device with square lattice connectivity $\mathcal{S}$ (see \Cref{fig:square}) comprising N qubits. Though it is simple to generalize to a rectangular shape, we assume a lattice of size $n \times n$, where $n=\sqrt{N}$, for simplicity. We sum the horizontal distance and the vertical distance between qubits with coordinates $(i, a)$ and $(j, b)$ where $i, j$ denote the horizontal coordinate and $a, b$ denote the vertical coordinate. The sum of all shortest paths is
\begin{align}
\begin{split}
    \sum L &= \sum_{a=1}^n \sum_{b=1}^n \sum_{j=1}^n \sum_{k=1}^n \left| k-j \right| + \left| b-a \right|\\
    &= 2 \sum_{a=1}^n \sum_{b=1}^n \sum_{j=1}^n \sum_{k=1}^n \left| k-j \right| \\
    &= 2 n^2 \sum_{j=1}^n \sum_{k=1}^n \left| k-j \right| \\
    &= 2 n^2 \frac{(n-1)n(n+1)}{3} \\
    &= 2 N \frac{(\sqrt{N}-1)\sqrt{N}(\sqrt{N}+1)}{3} \\
    &= 2 N \frac{\sqrt{N}(N-1)}{3} \, ,
\end{split}
\end{align}
so, according to \eqref{eq:L_general_2}, the average shortest path length is
\begin{equation}
    \langle L \rangle_\mathcal{S} = \frac{2\sqrt{N}}{3}
\end{equation}

\section{Curve fits} \label{app:curve-fits}

We present the numerical values of the parameters from the fit of the SWAP count data to the curve \eqref{eq:swap-curve} (setting $d=1$) for the path graph, for all the routing methods studied in the manuscript in \Cref{tab:swap-fit-path}. We also present the numerical values of the parameters from the fit of the depth data to the curve \eqref{eq:curve-depth-1d} for the \textit{basic} SABRE heuristics in \Cref{tab:depth-fit-path}.

We do the same for the parameters from the fits of the SWAP count and depth data to the curves \eqref{eq:swap-curve} (setting $d=2$) and \eqref{eq:curve-depth-2d} respectively, for the square connectivity, in \Cref{tab:swap-fit-square} and \Cref{tab:depth-fit-square}.

\section{Linear dependence of $S$ and $\tilde{D}$ on $G$} \label{app:g}

\begin{figure}
    \centering
    \includegraphics[width=0.9\columnwidth]{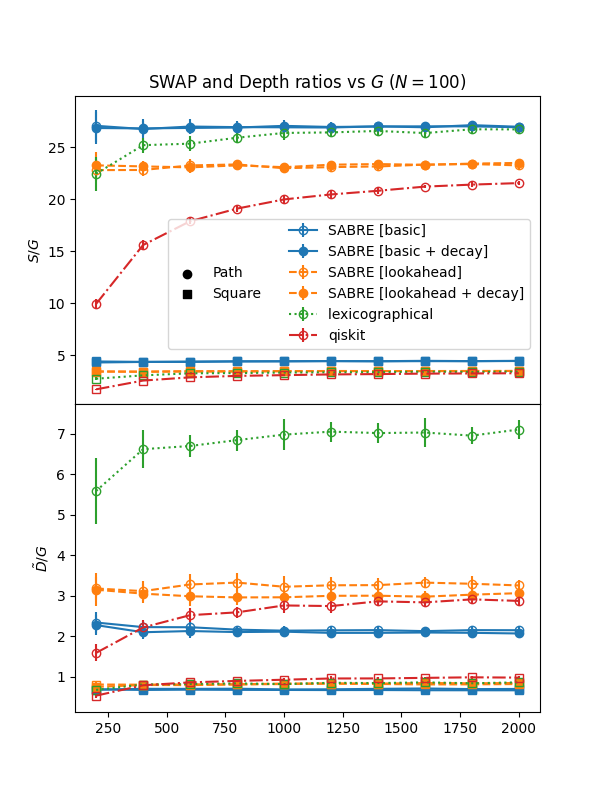} 
    \caption{SWAP gate count (top) and depth (bottom) ratios with respect to the number of gates of the original circuit, $G$, averaged over $50$ random circuits, for (a) the path connectivity, and (b) the square connectivity.}
    \label{fig:g}
\end{figure}

In \Cref{fig:g}, we plot the SWAP gate count $S$ and depth $\tilde{D}$, averaged over $50$ random circuits, and normalized by the number of gates $G$ of the original quantum circuit, for the path and square connectivities. We have chosen $N=100$. We observe that, for $G \gg N$, all the curves stabilize at a constant value for both the path and square connectivities for all the routing methods benchmarked, making it clear that the dependence of both quantities on $G$ is
\begin{align}
    S(G) & = \Theta(G) \, , \\
    \tilde{D}(G) & = \Theta(G) \, .
\end{align}

\section{Depth curve in one-dimensional devices} \label{app:depth-curve}

The term in parentheses in Eq.~\eqref{eq:curve-depth-1d} for one-dimensional connectivities serves the purpose of fixing a specific value for the offset of the curve, which must be fulfilled by construction, therefore reducing the number of parameters of the fit. It is chosen in such a way that the depth for the path connectivity is equal to $G$ for $N=2$, because for this given size, the connectivity does not require any SWAP gates, and all gates must be applied sequentially, therefore $\tilde{D}(2) = D(2) = G$.

Additionally, the parameter $A$ encapsulates the appropriate base of the logarithm, following from the identity $\log_b(a) = \frac{\log_x(a)}{\log_x(b)}$.

\end{document}